\documentclass[twocolumn,showpacs,prb,amsfonts,amsmath,amssymb,floatfix]{revtex4}
\usepackage{color}
\usepackage{hhline}
\usepackage{mathrsfs}
\usepackage{dcolumn}
\usepackage{multirow}
\usepackage{bm}
\usepackage{afterpage}
\usepackage{amsmath}
\usepackage{booktabs}
\usepackage{graphics}
\usepackage{graphicx}
\usepackage{hyperref}
\hypersetup{backref, colorlinks=true, linkcolor=blue, citecolor=blue}

\begin{document}

\title{Mechanism of  charge transfer/disproportionation  in $Ln$Cu$_{3}$Fe$_{4}$O$_{12}$ ($Ln$: Lanthanides)
}

\author{N.~Rezaei,$^{1,2}$ P.~Hansmann,$^{3}$~M.~S. Bahramy,$^{1,4}$ and~R.~Arita$^{1,4,5}$}
\affiliation{$^1$Quantum-Phase Electronics Center (QPEC) and Department of Applied Physics, The University of Tokyo,
             Hongo, Bunkyo-ku, Tokyo 113-8656, Japan\\
$^2$Department of Physics, Isfahan University of Technology, Isfahan, 84156-83111, Iran\\
$^3$Centre de Physique Th\'eorique, Ecole Polytechnique, CNRS-UMR7644, 91128 Palaiseau, France\\
$^4$RIKEN center for Emergent Matter Science (CEMS), Wako 351-0198, Japan\\
$^5$JST-PRESTO, Kawaguchi, Saitama 332-0012, Japan}

\date{\today}
\begin{abstract}
The Fe-Cu  intersite charge transfer and Fe charge disproportionation are  interesting phenomena observed in some $Ln$Cu$_3$Fe$_4$O$_{12}$ ($Ln$: Lanthanides) compounds containing light and heavy $Ln$ atoms, respectively.
We show that a change in the spin state is responsible for the intersite charge transfer in the light $Ln$ compounds. 
At the high spin state,  such systems prefer an unusual Cu-$d^8$ configuration, whereas at the low spin state they retreat to the normal Cu-$d^9$ configuration through a charge transfer from Fe  to Cu-$3d_{xy}$ orbital.
We  find that the strength of the crystal field splitting and the relative energy ordering between Cu-$3d_{xy}$ and
Fe-$3d$ states are the key parameters, determining the intersite charge transfer (charge disproportionation) in light (heavy) $Ln$  compounds. 
It is further proposed that the size of $Ln$ affects the onsite interaction strength of Cu-3$d$ states, leading to a strong modification of the Cu-$L_3$ edge spectrum, as observed by the X-ray absorption spectroscopy.
\end{abstract}

\maketitle
\section{Introduction}
A new class of double perovskites $A$Cu$_3$Fe$_4$O$_{12}$ ($A$CFO) has attracted
extensive attention in the recent years, as they exhibit unusual valence transitions, which have significant impact on their electrical and magnetic properties. 
{Such  valence transitions  can be induced by a small external stimulus rather than by carrier doping.}
{A prototypical example is the}
observed charge disproportionation between Fe atoms (2Fe$^{4+}\rightarrow$ Fe$^{3+}$+ Fe$^{5+}$)  
in CaCFO~\cite{CaCFO-exp1,CaCFO-exp2}, accompanied with  paramagnetic-ferrimagnetic and metal-insulator phase transitions.
The substitution of Sr for Ca turns the observed ferrimagnetic state in CaCFO into an antiferromagnetic (AFM) state~\cite{SrCFO-exp1}.
Remarkably, SrCFO exhibits a giant negative thermal expansion, ascribed to a temperature-induced  intersite charge transfer 
from Fe to Cu. 
Intriguingly, {the size of} $A^{2+}$ ion is the only relevant parameter, determining the type of the magnetic and valence transitions in these systems.\\
{Recently, very similar behaviors have also been  observed  in rather different $A$CFO systems with $A^{3+}$ ions such as 
La~\cite{LaCFO-exp1,LaCFO-exp2}, Bi~\cite{BiCFO-exp1,BiCFO-dft1} and Y~\cite{YCFO-exp1}. 
LaCFO and BiCFO  exhibit a temperature-induced intermetallic charge transfer 
(3Cu$^{3+}$+4Fe$^{3+}\rightarrow$ 3Cu$^{2+}$+4Fe$^{3.75+}$), giving rise to 
an insulator-metal transition as well as an abrupt volume reduction.
Such a sharp discontinuity in resistivity, susceptibility, and volume  implies a first-order phase transition in these systems.
Interestingly, Long $et~al.$~\cite{LaCFO-exp3} have recently discovered that applying pressure can also cause an intersite charge transfer in LaCFO.
On the other hand, YCFO exhibits 
charge disproportionation (8Fe$^{3.75+}\rightarrow$5Fe$^{3+}$+3Fe$^{5+}$) rather than
intersite charge transfer.}\\
{To understand the mechanism behind these phase transitions, Yamada $et~al.$~\cite{LnCFO}  recently synthesized a number of $Ln$CFO perovskites 
($Ln$ = Lanthanides: La, Pr, Nd, Sm, Eu, Gd, Tb, Dy, Ho, Er, Tm, Yb, Lu), and  systematically studied
their electronic and magnetic properties.
They  found that}
the compounds with light rare earth elements (LREE) exhibit intersite charge transfer with  
a temperature-induced paramagnetic-antiferromagnetic phase transition. The corresponding transition temperature was found to be higher for compounds with larger LREE.
In contrast, the compounds containing the heavy rare earth elements (HREE)
exhibited a charge disproportionation leading to a metal-semiconductor phase transition. Remarkably, the transition temperature in these systems was found to be rather insensitive to the size of $Ln$.
This study has accordingly raised the question how and why the size of  $Ln$  determines the type of the valence transition.\\
For LaCFO, it has been experimentally shown that the energy scale of the exchange coupling in the low temperature phase is
much larger than the transition temperature of the valence transition. This observation suggests that
the magnetic long-range order can not be the 
driving force for charge transfer~\cite{LaCFO-exp2}.
On the theory side, Li {\it et al.} performed a spin-density-functional calculation for LaCFO~\cite{LaCFO-dft1} 
and proposed a scenario in which the O-$2p$ orbitals mediate the Cu-Fe intermetallic charge transfer.
However, it is still unclear why the intersite charge transfer occurs only in LREE, and furthermore
why LaCFO has the highest transition temperature.
{The aim of the present work is to address these questions.
We  perform relaxed and fixed magnetic-moment calculations
for LaCFO and LuCFO; the two representative  systems exhibiting charge transfer and charge disproportionation, respectively. We show that the crystal field splitting and the relative ordering between Cu-$3d_{xy}$ and
Fe-$3d$ states are the key parameters, determining the type of valence transition}.\\ 

\section{Methods}
The total energy calculations were performed within the context of density functional theory using the Perdew-Burke-Ernzerhof exchange-correlation functional~\cite{PBE} and the projector-augmented-wave pseudopotentials~\cite{PAW} as implemented in the {\sc vasp} package~\cite{VASP}.
An energy cutoff of 600 eV was chosen for the expansion of the plane waves.
The Brillouin zone was sampled using a 11$\times$11$\times$11 Monkhorst-Pack $k$-mesh.
The Wannier functions (WFs) were constructed by  {\sc wannier90}  package~\cite{WANNIER,MLWF,Disentanglement-WF}.
The X-ray-absorption spectra calculations were performed within multiplet ligand field theory (MLFT) as implemented in the code of Ref.~\onlinecite{MLFT}.
In this method we obtain MLFT parameters from {\it ab initio} calculations unlike usual MLFT methods where these parameters are fit to experimental data.

\section{Results and Discussion}
\subsection{Electronic structure}
\begin{figure}[tp]
\begin{center}
\includegraphics[width=3.3in,clip]{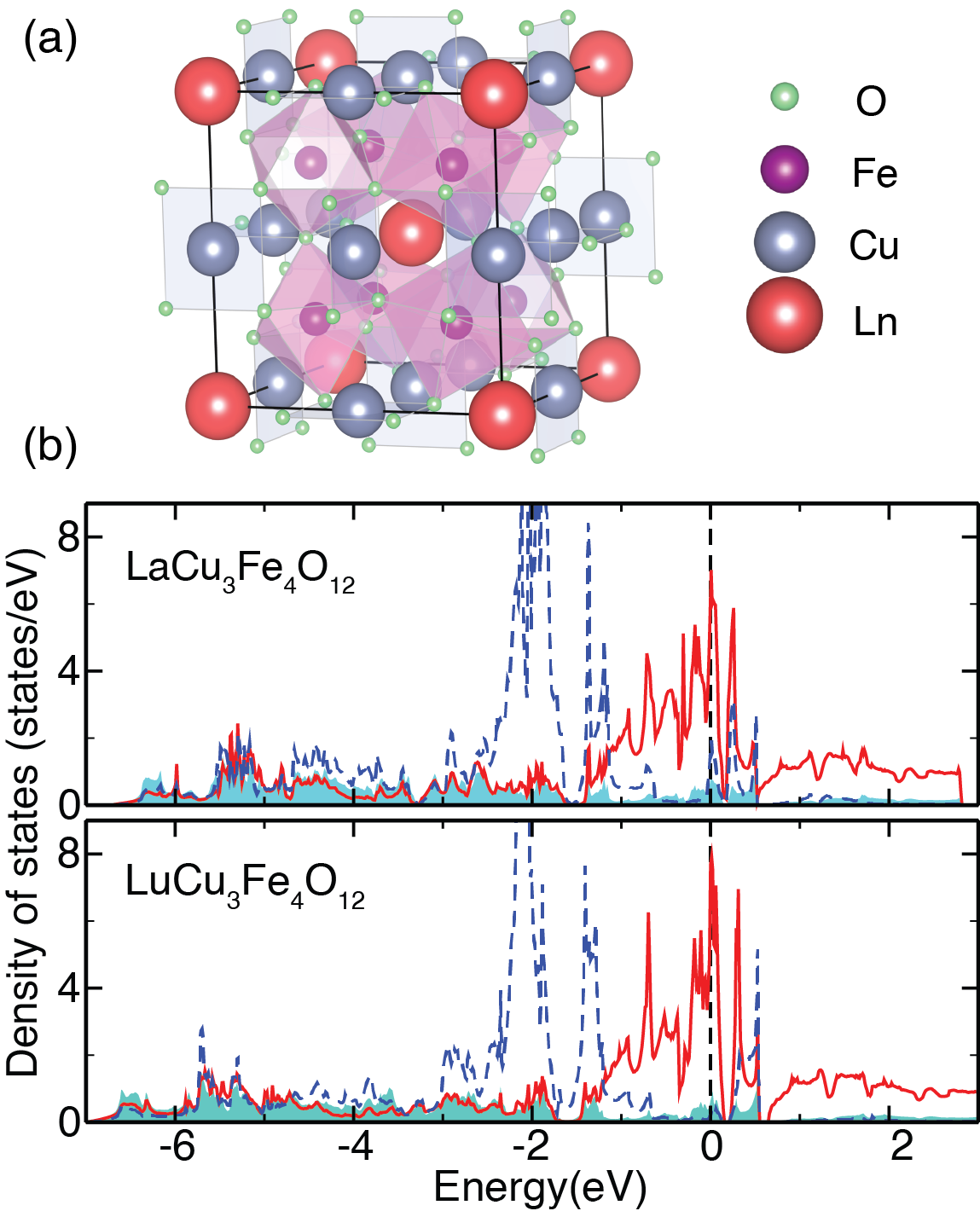}
\end{center}
\caption{(Color online). (a) Crystal structure of $A$-site-ordered $Ln$CFO ($Ln$: Lanthanides) double perovskites. (b) The corresponding partial density of states (PDOS) calculated for LaCFO and LuCFO. The PDOS corresponding to the Fe-$3d$, Cu-$3d$ and  O-$2p$ states are indicated by the solid (red) line, dashed (blue) line and shaded (blue) area, respectively. The Fermi energy is set to zero.}
\label{fig:Fig1}
\end{figure}
\begin{table}[bp,Table1]
\caption{\label{Tab1} Calculated crystal filed splitting ($\Delta_{t_{2g}-e_{g}}$) and the energy difference between the
 Fe-$t_{2g}$ and Cu-$3d_{xy}$ states($\Delta_{t_{2g}-Cu}$) for LaCFO and LuCFO.
}
\begin{ruledtabular}
\centering
\begin{tabular}{ccc}
          & $\Delta_{t_{2g}-e_{g}}$ (eV)& $\Delta_{t_{2g}-Cu}$ (eV) \\
\hline
LaCFO     & 1.79          & 0.44  \\
LuCFO     & 1.96          & 0.41  \\
\end{tabular}
\end{ruledtabular}
\end{table}
$Ln$CFO compounds crystallize in a cubic structure with $Im\bar{3}$ symmetry. As shown in Fig.~\ref{fig:Fig1}(a), the Fe atom is located at the center of a tilted FeO$_6$  octahedron, whereas the Cu atom is coordinated with four O atoms, together forming a square planar CuO$_4$ unit~\cite{LaCFO-exp4}.
In the present work, we have used the experimentally-identified structural parameters for both LaCFO and LuCFO at high temperature~\cite{LnCFO}.
In order to get insight into the character of the bands around the Fermi level, $E_F$,
we first study the electronic structure properties of LaCFO and LuCFO in the non-spin-polarized configuration.
Fig.~\ref{fig:Fig1}(b) shows the corresponding partial density of states (PDOS) for the two compounds.  
As can be seen, in this configuration, LaCFO  and LuCFO are both metallic, in agreement with the experiment. The octahedral crystal field acting on Fe splits its $3d$ orbitals into two manifolds $t_{2g}$ and $e_g$.
The three $t_{2g}$ orbitals span the energy region from -1.5 to 0.5 eV around  $E_F$ (which is set to zero).
The two Fe-$e_g$ orbitals are found to be above $E_F$.
Cu-$3d$ states, on the other hand, are almost fully occupied. The only exception is Cu-$3d_{xy}$, remaining partially unoccupied.
The PDOS in Fig.~\ref{fig:Fig1}(b) indicates a substantial overlap of Fe-$3d$, Cu-$3d$, and O-$2p$ states, therefore suggesting a strong hybridization between these states around $E_F$ in both LaCFO and LuCFO.
A tight-binding model can shed a light on the origin of the experimentally-observed different ground states
in LaCFO and LuCFO at  low temperatures. 
Considering the fact that the energy bands around $E_F$ are predominantly made of Fe-$3d$ and Cu-$3d_{xy}$, we have accordingly constructed a set of WFs for the respective conduction and valence bands using a basis set composed of these atomic orbitals.
For both LaCFO and LuCFO, the corresponding crystal field splitting ($\Delta_{t_{2g}-e_g}$) as well as the differential onsite energies between the Fe-$t_{2g}$ and Cu-$3d_{xy}$ orbitals ($\Delta_{t_{2g}-Cu}$) are summarized in table~\ref{Tab1}.
Clearly, $\Delta_{t_{2g}-e_g}$ in LuCFO is larger than that in LaCFO. This implies that Fe in LuCFO has less tendency to be in a $+3$ state. 
On the other hand, $\Delta_{t_{2g}-Cu}$ in LuCFO is smaller than that in LaCFO and, thus, less energy cost is required  to fill the Cu-$3d_{xy}$ states in LuCFO than it does in LaCFO.
As a result of the weak crystal filed splitting and large $\Delta_{t_{2g}-Cu}$, all Fe-$3d$ (Cu-$3d_{xy}$) orbitals  become fully occupied (unoccupied) in LaCFO. Accordingly, Fe goes to a Fe$^{+3}$ state and Cu stabilizes in an unusual Cu$^{+3}$ state.
\subsection{Intersite charge transfer and charge disproportionation}
\begin{figure}[htbp]
\begin{center}
\includegraphics[width=3.3in,clip]{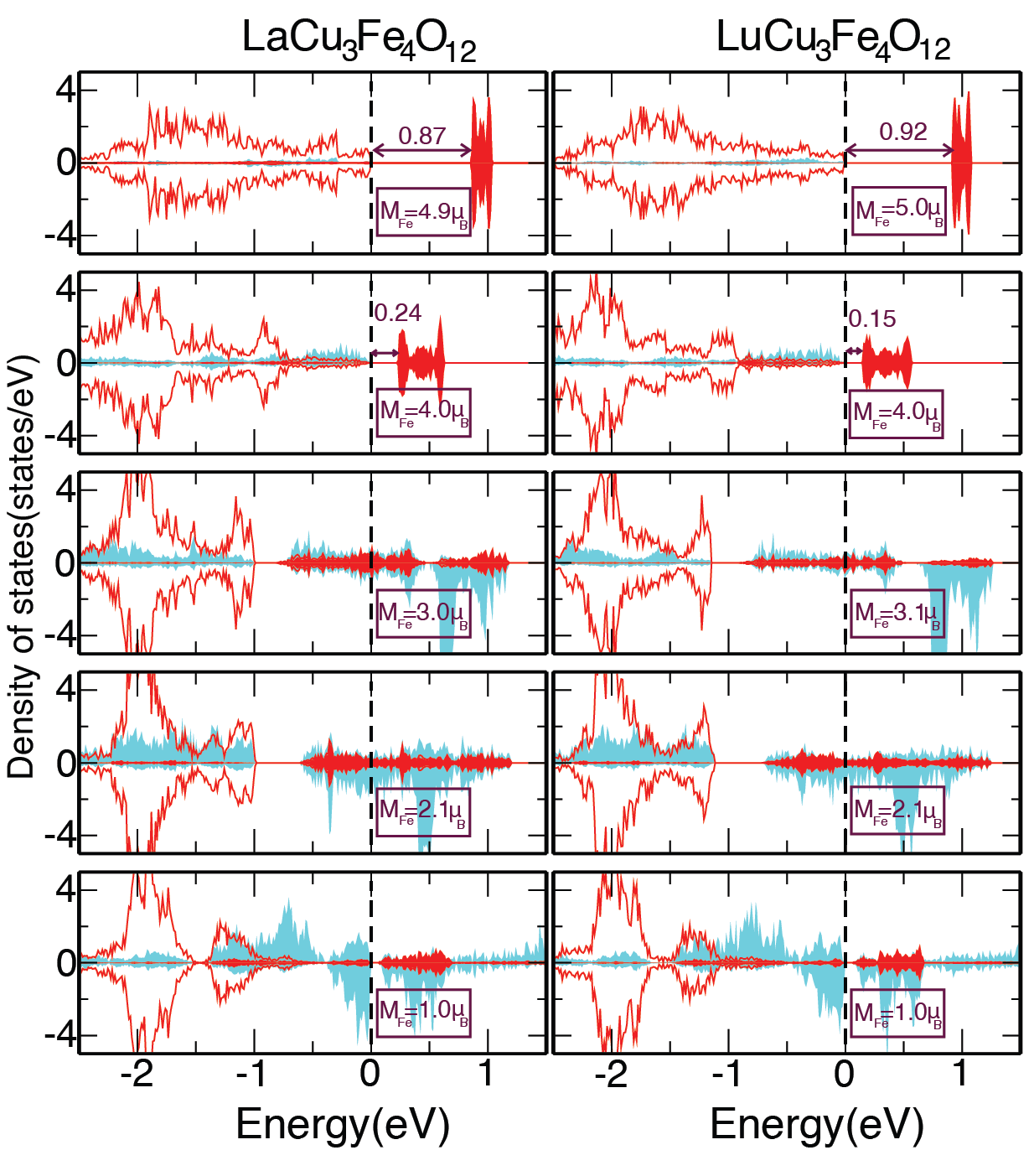}
\end{center}
\caption{(Color online). Partial density of states for Cu-$3d$ orbitals (red lines) and Fe-$3d$ orbitals (cyan shaded area, lightly shaded  in grayscale) as calculated under various constrained magnetizations. The red shaded area (darkly shaded in grayscale) indicates the contribution of Cu--$3d_{xy}$.}
\label{fig:Fig2}
\end{figure}

As mentioned earlier, upon increasing temperature an intersite charge transfer accompanied with the antiferromagnetic-paramagnetic phase transition occurs in LaCFO.
In order to understand the mechanism of intersite charge transfer, we examine the evolution of electronic structure of LaCFO from an antiferromagnetic state to a non-magnetic state by constraining the size and direction of the magnetic moment on Fe.
To be consistent with the experiment, we adopt the G-type AFM spin arrangement for  the Fe atoms.
We start from the fully spin-polarized configuration ($M_{\text {Fe}}$~=~5.0$\mu_{B}$) and examine how the density 
of states evolves by decreasing $M_{\text {Fe}}$.
The calculated PDOS for LaCFO indicates that this system is an insulator with a band gap changing from 0.87 to 0.24 eV, when reducing $M_{\text {Fe}}$ from 4.9  to 4.0$\mu_{B}$ (see Fig.~\ref{fig:Fig2}). This qualitatively agrees with the experimental finding that LaCFO in the AFM phase behaves like an insulator.
As can be seen in Fig.~\ref{fig:Fig2}, at high magnetic states, while the PDOS of Fe-$3d$ are either fully (for $M_{\text {Fe}}$~=~4.9$\mu_{B}$) or predominantly (for $M_{\text {Fe}}$~=~4.0$\mu_{B}$) below $E_F$, the corresponding Cu-$3d_{xy}$ states appear above $E_F$.
For both the Cu-$3d_{xy}$ and Fe-$3d$, the position of the respective PDOS is quite sensitive to the value of $M_{\text {Fe}}$.
Upon decreasing $M_{\text {Fe}}$, the Cu-$3d_{xy}$  (Fe-$3d$) states shift to lower (higher) energies,
and eventually cross $E_F$ when $M_{\text {Fe}}$ becomes 3.0$\mu_{B}$ or less (see Fig.~\ref{fig:Fig2}).
\begin{figure}[htbp]
\begin{center}
\includegraphics[width=3.3in,clip]{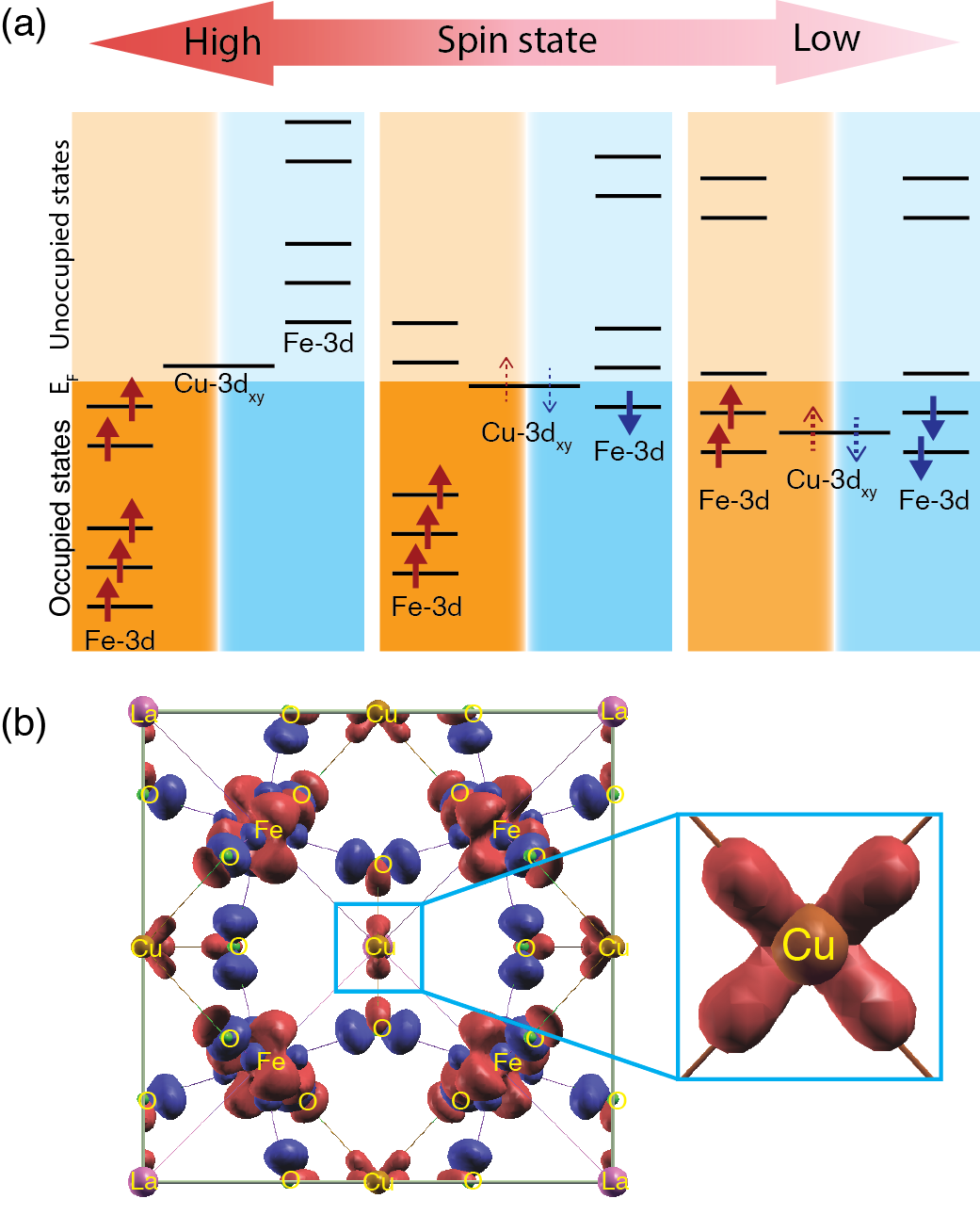}
\end{center}
\caption{(Color online). (a) Schematic diagram of intersite charge transfer in $Ln$CFO compounds. (b)  The excess and depletion charge distributions obtained from subtracting the total charge density of LaCFO in  $M_{\text {Fe}}$=4.9$\mu_{B}$ from that in $M_{\text {Fe}}$=2.1$\mu_{B}$.
The excess (depletion)  charges are indicated in red (blue) color. The  inset shows a magnified view of the excess and depletion charge distributions around the central Cu atom.}
\label{fig:Fig3}
\end{figure}
As a result, at low magnetic states, the Cu-$3d_{xy}$ (Fe-$3d$) states become mainly occupied (unoccupied), as schematically depicted in Fig.~\ref{fig:Fig3}(a). It is worth noting that, the O-$2p$ orbitals play an important role in this process, as proposed in Ref.~\onlinecite{LaCFO-dft1}. To elucidate this, in Fig.~\ref{fig:Fig3}(b) we show the excess and depletion charge distributions obtained from subtracting the total charge density of LaCFO with  $M_{\text {Fe}}$~=~4.9$\mu_{B}$ from that with $M_{\text {Fe}}$~=~2.1$\mu_{B}$. The Cu atoms clearly gain a pure excess charge, spatially distributed in their $3d_{xy}$ orbital. On the other hand, the Fe atoms undergo a partial charge loss, as expected. This charge loss is obviously compensated by the excess charge on Cu atoms, but importantly mediated through the O atoms. To be more precise, by lowering $M_{\text {Fe}}$, the charge density around O atoms is redistributed such that Cu-O (Fe-O) bonds become more (less) electron-rich. This indicates that the $d^8$ configuration of copper in the AFM state transforms into a $d^9$ configuration in the non-magnetic phase. Consequently the intersite charge transfer observed in LaCFO (and the other LREE compounds) is expected to be magnetically driven.
\begin{figure}[ht]
\begin{center}
\includegraphics[width=3.3in,clip]{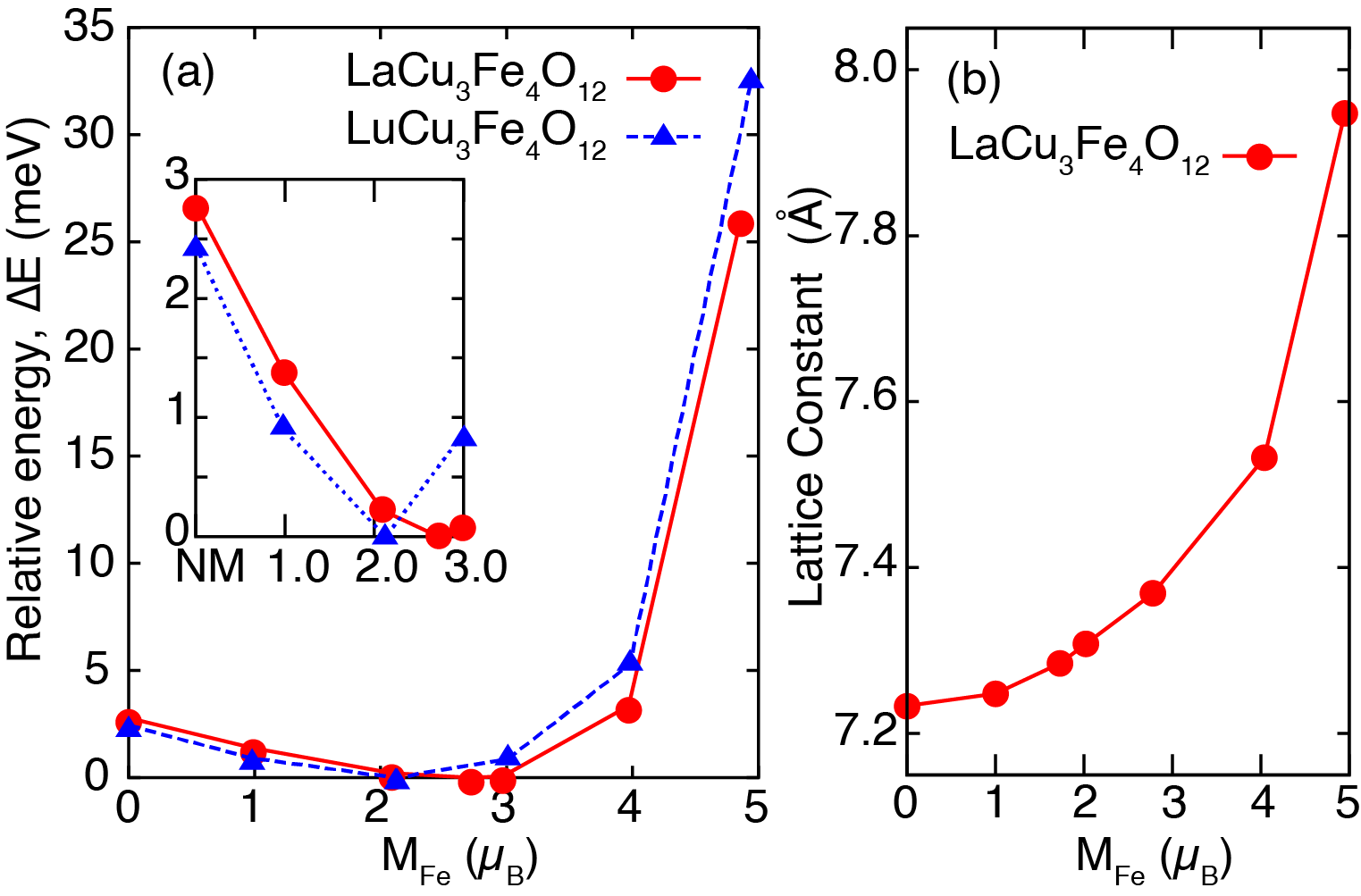}
\end{center}
\caption{(Color online). (a) Relative energy($\Delta E$) between the magnetically constrained configurations and the self-consistently obtained G-type antiferromagnetic configuration
as a function of the absolute value of $M_{\text {Fe}}$ calcualted  for LaCFO(triangles) and LuCFO(circles).The inset shows a magnified view of $\Delta E$ for $M_{\text {Fe}}\le$3.0$\mu_{B}$.
(b) Optimized lattice constant as a function of absolute value of $M_{\text {Fe}}$ for LaCFO.}
\label{fig:Fig4}
\end{figure}
We have also performed the constrained magnetization calculations for LuCFO (see Fig.~\ref{fig:Fig2}). It can be seen that when
$M_{\text {Fe}}<$~4.0$\mu_{B}$, the Cu-$3d_{xy}$ orbital becomes partially filled.
The comparison of the $M_{\text {Fe}}$ dependence of the total energy reveals why LaCFO prefers the ground state LaCu$^{+3}_3$Fe$^{+3}_{4}$O$_{12}$,
whereas LuCFO prefers  LuCu$^{+3}_3$Fe$^{+3}_{5/2}$Fe$^{+5}_{3/2}$O$_{12}$ at low temperature.
Fig.~\ref{fig:Fig4}(a) shows the relative energy ($\Delta E$) between the magnetically constrained configurations and the self-consistently obtained G-type AFM configuration for the both systems.
It is clear that in comparison with LaCFO,  LuCFO requires increasingly higher energy cost to be driven away from its self-consistent configuration into a configuration with larger $M_{\text {Fe}}$. This accordingly means that Fe in LuCFO has much less tendency to be in the Fe$^{+3}$ state than it does in LaCFO. This situation completely changes when $M_{\text {Fe}}$ is lowered from its self-consistent value. In this case, the calculated $\Delta E$ is mildly sensitive to the value of $M_{\text {Fe}}$ and its seems to be relatively smaller for LuCFO than for LaCFO.
This indicates that  Cu$^{+2}$ can be easily formed in LuCFO. Considering the facts that Cu normally prefers to be in Cu$^{+2}$ state and Fe generally disfavors Fe$^{+4}$ state, it becomes clear why LuCFO behaves as if it has fewer Fe$^{+4}$ ions. On this basis, we can further expect that as the temperature decreases the Cu$^{+2}$ state
remains unaffected, whereas the Fe$^{+3.75}$ ions transform to Fe$^{+3}$ and Fe$^{+5}$ states.
In contrast, LaCFO requires less energy cost to be in high spin state. Therefore, it prefers to transfer charge from its Cu to Fe atoms and form Fe$^{+3}$ (3Cu$^{2+}$+4Fe$^{3.75+}\rightarrow$ 3Cu$^{3+}$+4Fe$^{+3}$).
A noteworthy point is how the lattice constant is affected by the changes in magnetization.
For this purpose, we first fixed the position of Fe at its symmetric point, then
relaxed the position of O, and eventually optimized the lattice parameter. According to these calculations, when $M_{\text {Fe}}$~=~4.9$\mu_B$,  the lattice constant reaches its maximum (see Fig.~\ref{fig:Fig4}(b)). This behavior is consistent
with the experimentally-observed volume expansion in the AFM phase.
 \subsection{X-ray absorption spectra}
Yamada $et~al.$~\cite{LnCFO}  have also reported X-ray absorption spectra (XAS) for the LREE group. According to their measurements, at low temperatures the  Cu-$L_3$ edge splits into  two peak structures, which is an indication that the ground state of Cu has $d^8$ configuration~\cite{XAS-CuL3}. The relative intensity of these two peaks has been observed to be dependent on the size of $Ln$.
While it is higher for the higher energy peak in the case of $Ln$=Nd, it turns out to be lower for the same peak for $Ln$=Tb.
To examine whether our calculation can reproduce this finding, we have simulated the Cu-$L_3$ absorption edge using the MLFT method~\cite{MLFT}. 
 We consider a cluster composed of  Cu-$3d$ and the core shell Cu-$2p$ states. We also include O-$2p$ ligand states  ``$L$'', but only those which have a finite overlap with the Cu-$3d$ states, as extracted from the {\it ab initio} calculations. 
Due to the fact that spin-orbit coupling is small, S$_z$ is \textit{approximately} a good quantum number, and its expectation value for the ground state would be 0, i.e. a spin singlet.
  The higher order $3d$ Slater integrals F$^2$ and F$^4$, the interaction between the core-hole and valence electrons as well as the core-hole spin-orbit coupling have been deduced from the atomic Hartree Fock values while
the charge transfer energy $\Delta$ and the Hubbard U are, as usual, taken as adjustable parameters in the calculation~\cite{Parameters,core-hole-note}.

\begin{figure}[tp]
\begin{center}
\includegraphics[width=3.4in,clip]{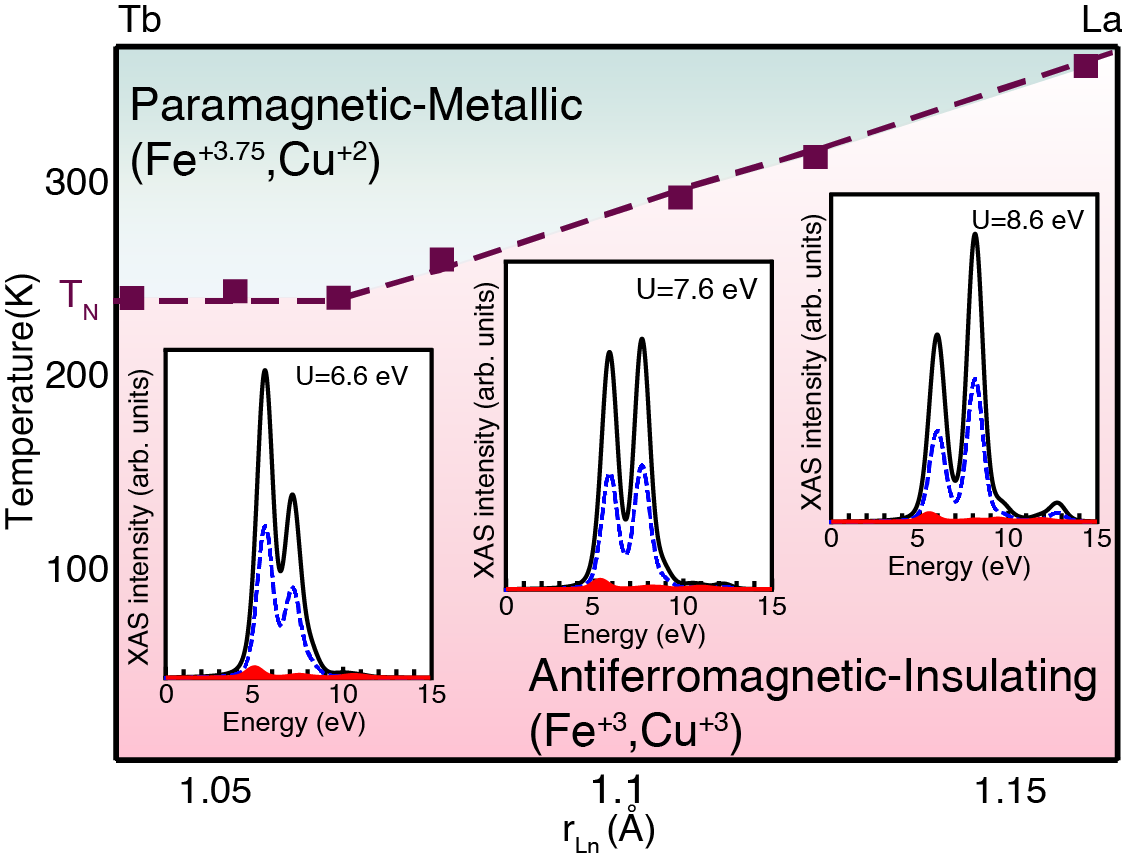}
\end{center}
\caption{(Color online). $L_3$ X-ray absorption edge of Cu  calculated for  {$Ln$CFO  with  different values  for the Cu onsite interaction parameter $U$.  The incident light is assumed to be either non-polarized  (black solid line) or polarized parallel  (blue dashed line) or perpendicular (red  shaded area) to the crystalline $ab$-plane.}
The background phase diagram indicates the relation between the magnetic phase transition with the size ($r_{Ln}$) of the $Ln$ atom in $Ln$CFO. $T_N$ denotes the Neel temperature. }
\label{fig:Fig5}
\end{figure}

Setting $\Delta$ and $U$ to the typical values for transition metal oxides, we can successfully reproduce the experimentally-observed spectra with the pronounced two-peak structure in the Cu-$L_3$ edge. As shown in Fig.~\ref{fig:Fig5},  almost the entire spectral weight comes from 
the absorption of incident lights, polarized in the $xy$ plane.
This reflects the fact that on each copper site the Cu-$3d_{xy}$ orbital (as defined in the local coordinate system)  
is the only possible final state for the excited core electrons~\cite{xas-note1}.
The shape of the relative intensity of the peaks strongly depends on the value of $U$ but rather insensitive to the choice of $\Delta$. 
Note that  the Hubbard U parameter is taken as an adjustable parameter in our calculations. However,  as a cross-check we have separately estimated the possible range of U for Cu-3d states in LaCFO by performing a set of  constrained local density approximation calculations within both the full-potential\cite{wien-U} and pseudo-potential\cite{pwscf-U} formalisms.  The respective value of U is accordingly found to be in the range of 7.81 eV (obtained from the full-potential calculations)  to 10.43 eV (obtained from the pseudo-potential calculations). This range obviously includes U = 8.6 eV, the value that we have chosen for LaCFO in our XAS  calculations. 
 As can be seen, by increasing $U$, the spectral weight transfers monotonously from the lower energy peak to the higher one.
This is due to the fact that the ground state can be excited into two different excited states, namely $|1\rangle\equiv|\text{core-hole}~ 3d^9 L^{10}\rangle$ and $|2\rangle\equiv|\text{core-hole} ~3d^{10} L^9\rangle$ (where $L$ denotes the hybridizing ligand states as described above). The observed splitting of the $L_3$ edge originates from the core-hole valence electron interaction. Increasing $U$ increases the energy of placing the excited core electron into the $d$-shell. Hence, the related processes (leading to final state $|2\rangle$) are energetically shifted up leading to the observed spectral weight transfer~\cite{xas-note2}. 
This effect allows us to exploit the spectral structure as an experimental probe for the onsite interaction strength $U$.
In fact, when we revisit the experimental spectra for the different $Ln$CFO materials in this light,
we see that an increase in the onsite interaction should be expected for larger $Ln$ atoms,
consistent with  more localized Cu-$3d$ orbitals and a higher T$_N$, as depicted by the phase diagram shown in Fig.~\ref{fig:Fig5}.\\
\section{Conclusion}
In conclusion, we  studied the mechanism of charge transfer/disproportionation in the double perovskite transition metal oxides $Ln$Cu$_3$Fe$_4$O$_{12}$.  
Examining their electronic structure at various constrained magnetic configurations, we shed light on the origin of the observed magnetic and valence transitions in these systems.   
It was further explained, why the size of the rare earth element can be the main parameter determining the type of valence transition in such systems. 
We also calculated the X-ray absorption spectra for $Ln$CFO compounds and confirmed the presence of the experimentally-reported  two peak structure at Cu-$L_3$-edge. The relative intensity of these two peaks were shown to be sensitive to the strength (and hence applicable as a probe) of the onsite interaction.
\begin{acknowledgments}
We thank I. Yamada for providing us  the preprint of Ref.~\onlinecite{LnCFO} in prior to its publication.  This work was supported  by the Japan Society for the Promotion of Science (JSPS) through the "Funding Program for World-Leading Innovative R\&D on Science and Technology (FIRST Program)", initiated by the council for Science and Technology Policy (CSTP), JST-PRESTO and Isfahan University of Technology. 
\end{acknowledgments}


\begin{thebibliography}{999}
\bibitem{core-hole-note}{A reduction of the Hartree Fock values for the core hole valence interaction leads to: i) a reduction of the splitting of the two peaks in the L3 edge which is smaller than what is seen in experiment and ii) that, in any case, the trend that we see in the ratio of the peaks remains unchanged. Due to the fact that the final state exciton is a very localized - atomic like - entity it comes with no surprise that the unscreened Hartree Fock  parameters for this interaction (which we have employed) yield the best agreement with experiment.}
\bibitem{xas-note1}{Due to the fact that the absorption spectrum represents an average over all Cu 
sites in the unit cell, this polarization effect could only be observed in a more sophisticated experimental set-up such as what used for the
partial fluorescence yield measurements. }
\bibitem{xas-note2}{Indeed, further increasing of U would leads to a complete separation of spectra associated with either $|1\rangle$ or $|2\rangle$.}
\end{thebibliography}
\end{document}